\documentclass[10pt,conference]{IEEEtran}
\IEEEoverridecommandlockouts
\usepackage{amsmath,amssymb,amsthm,booktabs,tikz,pgfplots}
\usetikzlibrary{arrows.meta,positioning,calc}
\usepgfplotslibrary{fillbetween}
\pgfplotsset{compat=1.18}
\usepackage{cite}
\usepackage{acronym}
\usepackage[hidelinks]{hyperref}
\usepackage{etoolbox}
\usepackage[normalem]{ulem}
\newif\ifshowrevisionmarks
\showrevisionmarksfalse

\newcommand{\commentadd}[1]{#1}
\newcommand{\reviewfocus}[1]{#1}
\makeatletter
\def\@IEEENORMtitlevspace{0.6\baselineskip}
\def\@IEEEMINtitlevspace{0.4\baselineskip}
\apptocmd{\normalsize}{%
 \abovedisplayskip=4.5pt plus 2pt minus 1pt
 \belowdisplayskip=4.5pt plus 2pt minus 1pt
 \abovedisplayshortskip=2pt plus 1pt
 \belowdisplayshortskip=3pt plus 1pt minus 1pt
}{}{}
\patchcmd{\section}{1.5ex plus 1.5ex minus 0.5ex}
 {1.1ex plus 0.5ex minus 0.2ex}{}{}
\patchcmd{\section}{0.7ex plus 1ex minus 0ex}
 {0.5ex plus 0.3ex minus 0ex}{}{}
\patchcmd{\subsection}{1.5ex plus 1.5ex minus 0.5ex}
 {1.1ex plus 0.5ex minus 0.2ex}{}{}
\patchcmd{\subsection}{0.7ex plus .5ex minus 0ex}
 {0.5ex plus 0.3ex minus 0ex}{}{}
\makeatother
\AtBeginDocument{\normalsize}
\acrodef{isac}[ISAC]{integrated sensing and communication}
\acrodef{bs}[BS]{base station}
\acrodef{ue}[UE]{user equipment}
\acrodef{los}[LoS]{line-of-sight}
\acrodef{snr}[SNR]{signal-to-noise ratio}
\acrodef{cdf}[CDF]{cumulative distribution function}
\acrodef{crb}[CRB]{Cram\'er--Rao bound}
\acrodef{qcd}[QCD]{quickest change detection}
\acrodef{kl}[KL]{Kullback--Leibler}
\acrodef{ula}[ULA]{uniform linear array}
\acrodef{tv}[TV]{total variation}
\newtheorem{theorem}{Theorem}
\newtheorem{corollary}{Corollary}
\newtheorem{assumption}{Assumption}
\newcommand{\E}{\mathbb E}
\newcommand{\Prb}{\mathbb P}
\newcommand{\CN}{\mathcal{CN}}
\DeclareMathOperator{\logit}{logit}
\definecolor{monitor}{RGB}{0,112,139}
\definecolor{comm}{RGB}{180,91,36}
\title{Current--Future Rate Tradeoffs in ISAC\\Under Random Path Changes}
\author{
\IEEEauthorblockN{Hang Ruan\IEEEauthorrefmark{1},
Zhaofeng Liu\IEEEauthorrefmark{2}, and
Taneli Riihonen\IEEEauthorrefmark{1}}
\IEEEauthorblockA{\IEEEauthorrefmark{1}Faculty of Information Technology and Communication Sciences,
Tampere University, Finland\\
\IEEEauthorrefmark{2}Northeastern University, USA\\
\{hang.ruan, taneli.riihonen\}@tuni.fi; zhaofeliu3-c@my.cityu.edu.hk}
\thanks{This work was supported by the European Union's Marie
Sk\l{}odowska-Curie Actions (MSCA) Postdoctoral Fellowship project
ISAC-SS, Grant Agreement No. 101211218 and in part by the Research
Council of Finland.}
}

\def\asymSweepAOne{(0.01,0.656659497701) (0.015,0.629191671805) (0.02,0.602367974599) (0.025,0.576263861868) (0.03,0.551277924938) (0.035,0.527017969568) (0.04,0.503270879133) (0.045,0.480032521016) (0.05,0.457299151625) (0.055,0.435067553054) (0.06,0.413335240947) (0.065,0.392100795067) (0.07,0.371364420152) (0.075,0.351128989578) (0.08,0.331402263865) (0.085,0.312202691111) (0.09,0.293581934585) (0.095,0.275852499658) (0.1,0.258823895067) (0.105,0.242229166875) (0.11,0.226070984169) (0.115,0.210515687521) (0.12,0.199085682812) (0.125,0.188093810408) (0.13,0.177561480656) (0.135,0.16734529964) (0.14,0.157437749185) (0.145,0.147843779645) (0.15,0.138572581461) (0.155,0.1296417606) (0.16,0.121089627631) (0.165,0.112994063594) (0.17,0.105206923544) (0.175,0.0976842613384) (0.18,0.0904434170548) (0.185,0.0835282448339) (0.19,0.0769900493475) (0.195,0.0706288687582) (0.2,0.0644142410578) (0.205,0.0583388504462) (0.21,0.0523949699418) (0.215,0.0481177126486) (0.22,0.0449194073719) (0.225,0.0417707637281) (0.23,0.0386650248516) (0.235,0.0355950467666) (0.24,0.032553596032) (0.245,0.0295340397616) (0.25,0.026531747452) (0.255,0.0235466338758) (0.26,0.0205870660734) (0.265,0.0176739850332) (0.27,0.0148405804923) (0.275,0.0121208709687) (0.28,0.00953259621187) (0.285,0.00705087042207) (0.29,0.0045331515269) (0.295,0.00190147467576) (0.3,0) (0.305,0) (0.31,0) (0.315,0) (0.32,0) (0.325,0) (0.33,0) (0.335,0) (0.34,0) (0.345,0) (0.35,0) (0.355,0) (0.36,0) (0.365,0) (0.37,0) (0.375,0) (0.38,0) (0.385,0) (0.39,0) (0.395,0.00155833475472) (0.4,0.00384419392037) (0.405,0.00627234242446) (0.41,0.00882832509335) (0.415,0.0114878432831) (0.42,0.0142318048124) (0.425,0.0170467813625) (0.43,0.0199246319302) (0.435,0.0228614954258) (0.44,0.0258565754861) (0.445,0.0289110825336) (0.45,0.0320274523491)}
\def\asymSweepATwo{(0.01,0.312311152042) (0.015,0.296872558864) (0.02,0.281826866574) (0.025,0.267166970108) (0.03,0.252868127496) (0.035,0.238909611784) (0.04,0.225291031439) (0.045,0.212013029083) (0.05,0.199076533601) (0.055,0.186482773503) (0.06,0.174233294867) (0.065,0.162329978094) (0.07,0.150775034727) (0.075,0.139570932073) (0.08,0.128720098702) (0.085,0.118223952972) (0.09,0.108079495912) (0.095,0.101205769366) (0.1,0.0950830217792) (0.105,0.0891643350321) (0.11,0.0834454811362) (0.115,0.0779212615017) (0.12,0.0725855260348) (0.125,0.0674318340397) (0.13,0.0624555690376) (0.135,0.057654310292) (0.14,0.0530252921727) (0.145,0.0485645896995) (0.15,0.0442664993301) (0.155,0.0401227258209) (0.16,0.0361217679287) (0.165,0.0336807059365) (0.17,0.0319605000595) (0.175,0.0302746958717) (0.18,0.0286201513594) (0.185,0.0269938272779) (0.19,0.0253924988268) (0.195,0.0238126157905) (0.2,0.0222502569311) (0.205,0.0207011056201) (0.21,0.0191604463911) (0.215,0.0176231978) (0.22,0.0160840072679) (0.225,0.0145374571488) (0.23,0.0129784884836) (0.235,0.0114032720787) (0.24,0.00981094403285) (0.245,0.00820664017498) (0.25,0.00660524620985) (0.255,0.00503204381359) (0.26,0.00351348128969) (0.265,0.00205916846354) (0.27,0.000651408787919) (0.275,0) (0.28,0) (0.285,0) (0.29,0) (0.295,0) (0.3,0) (0.305,0) (0.31,0) (0.315,0) (0.32,0) (0.325,0) (0.33,0) (0.335,0) (0.34,0) (0.345,0) (0.35,0) (0.355,0) (0.36,0) (0.365,0) (0.37,0) (0.375,0) (0.38,0) (0.385,0) (0.39,0) (0.395,0) (0.4,0) (0.405,0) (0.41,0) (0.415,0) (0.42,0) (0.425,0) (0.43,0) (0.435,0) (0.44,0) (0.445,0) (0.45,0)}
\def\asymSweepBOne{(0.01,0.61861614098) (0.015,0.592811134094) (0.02,0.567595707072) (0.025,0.543037536335) (0.03,0.519473691219) (0.035,0.496558402393) (0.04,0.474112843212) (0.045,0.452133530907) (0.05,0.430617328269) (0.055,0.409561546118) (0.06,0.388964117521) (0.065,0.368823895838) (0.07,0.34914117984) (0.075,0.329918699336) (0.08,0.311163688152) (0.085,0.292893198868) (0.09,0.275154280374) (0.095,0.258215343144) (0.1,0.241902725583) (0.105,0.225992184738) (0.11,0.210488661381) (0.115,0.195402690678) (0.12,0.184287438865) (0.125,0.173943647962) (0.13,0.164080174505) (0.135,0.154513307514) (0.14,0.145233839961) (0.145,0.136248360089) (0.15,0.127566711216) (0.155,0.119205379939) (0.16,0.111197141194) (0.165,0.103601058027) (0.17,0.0962980555482) (0.175,0.0892596014623) (0.18,0.0825088626856) (0.185,0.0761037761158) (0.19,0.0701247803594) (0.195,0.0643203001824) (0.2,0.0586520201797) (0.205,0.0531131358529) (0.21,0.0476963413466) (0.215,0.0440001309819) (0.22,0.041264498163) (0.225,0.0385701850648) (0.23,0.0359094850765) (0.235,0.0332734185212) (0.24,0.0306516011305) (0.245,0.0280323866585) (0.25,0.025403738668) (0.255,0.0227557827059) (0.26,0.0200866870165) (0.265,0.0174133942875) (0.27,0.0147836003585) (0.275,0.0122703616007) (0.28,0.00993292301541) (0.285,0.00776628771251) (0.29,0.00558315400375) (0.295,0.00326393909316) (0.3,0.000749421293333) (0.305,0) (0.31,0) (0.315,0) (0.32,0) (0.325,0) (0.33,0) (0.335,0) (0.34,0) (0.345,0) (0.35,0) (0.355,0) (0.36,0) (0.365,0) (0.37,0) (0.375,0) (0.38,0) (0.385,0) (0.39,0) (0.395,0) (0.4,8.12556209016e-05) (0.405,0.00216795857535) (0.41,0.00435924991199) (0.415,0.00663072959142) (0.42,0.00896595688362) (0.425,0.0113556256374) (0.43,0.0137954986188) (0.435,0.0162846431363) (0.44,0.0188241435045) (0.445,0.0214162362573) (0.45,0.0240637647611)}
\def\asymSweepBTwo{(0.01,0.248910937911) (0.015,0.235761361519) (0.02,0.22291268161) (0.025,0.210357778878) (0.03,0.198078634212) (0.035,0.186062431737) (0.04,0.174309963464) (0.045,0.162823287449) (0.05,0.151605325532) (0.055,0.14065986031) (0.06,0.129991487909) (0.065,0.119605517283) (0.07,0.109507799882) (0.075,0.0997044519073) (0.08,0.0902013643854) (0.085,0.0810031763088) (0.09,0.0736836096033) (0.095,0.0684789359079) (0.1,0.0634632960244) (0.105,0.0586353767014) (0.11,0.0539920844802) (0.115,0.0495282539852) (0.12,0.0452364988699) (0.125,0.0411080137047) (0.13,0.0371365786607) (0.135,0.0333206919138) (0.14,0.0296592053135) (0.145,0.0261501810843) (0.15,0.0227904925985) (0.155,0.0204315249192) (0.16,0.0193890961053) (0.165,0.0183836903968) (0.17,0.0174102448431) (0.175,0.0164648195957) (0.18,0.0155447431819) (0.185,0.0146482111087) (0.19,0.013773559476) (0.195,0.0129188170759) (0.2,0.0120815336114) (0.205,0.0112586171834) (0.21,0.0104461326505) (0.215,0.00963906455583) (0.22,0.00883106317792) (0.225,0.00801421419397) (0.23,0.00717889942357) (0.235,0.00631385073042) (0.24,0.0054066138825) (0.245,0.00444518305285) (0.25,0.00342331445272) (0.255,0.00235422183407) (0.26,0.00128970550873) (0.265,0.00031055991861) (0.27,0) (0.275,0) (0.28,0) (0.285,0) (0.29,0) (0.295,0) (0.3,0) (0.305,0) (0.31,0) (0.315,0) (0.32,0) (0.325,0) (0.33,0) (0.335,0) (0.34,0) (0.345,0) (0.35,0) (0.355,0) (0.36,0) (0.365,0) (0.37,0) (0.375,0) (0.38,0) (0.385,0) (0.39,0) (0.395,0) (0.4,0) (0.405,0) (0.41,0) (0.415,0) (0.42,0) (0.425,0) (0.43,0) (0.435,0) (0.44,0) (0.445,0) (0.45,0)}
\def\asymSweepCOne{(0.01,0.557595264049) (0.015,0.53418231073) (0.02,0.511285539129) (0.025,0.488964601819) (0.03,0.467501238677) (0.035,0.44659428025) (0.04,0.426095436631) (0.045,0.406002232417) (0.05,0.386312639989) (0.055,0.367025109214) (0.06,0.348138645252) (0.065,0.329652988298) (0.07,0.311569000007) (0.075,0.293889482483) (0.08,0.276621007872) (0.085,0.259778680677) (0.09,0.243403984283) (0.095,0.227723866756) (0.1,0.212581237584) (0.105,0.197788739196) (0.11,0.183352247062) (0.115,0.169952669059) (0.12,0.16011735235) (0.125,0.150703544586) (0.13,0.141778550085) (0.135,0.133112765689) (0.14,0.124696446972) (0.145,0.116540463887) (0.15,0.108658086185) (0.155,0.101067311595) (0.16,0.09379805142) (0.165,0.0868961664312) (0.17,0.0802690780296) (0.175,0.0739015960962) (0.18,0.067823259206) (0.185,0.062107782589) (0.19,0.056872920827) (0.195,0.0518089346671) (0.2,0.0468682301468) (0.205,0.0420452093552) (0.21,0.0378916969275) (0.215,0.035691925082) (0.22,0.0335348474175) (0.225,0.0314149905849) (0.23,0.0293256058011) (0.235,0.0272581804624) (0.24,0.0252018024609) (0.245,0.0231424286146) (0.25,0.0210623328048) (0.255,0.0189406536714) (0.26,0.0167575801684) (0.265,0.0145077628426) (0.27,0.0122292672224) (0.275,0.0100316955258) (0.28,0.00805284989374) (0.285,0.00634306375859) (0.29,0.0046795121103) (0.295,0.00290561391345) (0.3,0.000952952136252) (0.305,0) (0.31,0) (0.315,0) (0.32,0) (0.325,0) (0.33,0) (0.335,0) (0.34,0) (0.345,0) (0.35,0) (0.355,0) (0.36,0) (0.365,0) (0.37,0) (0.375,0) (0.38,0) (0.385,0) (0.39,0) (0.395,0) (0.4,0) (0.405,0) (0.41,0.000253808120557) (0.415,0.00208094018866) (0.42,0.00394194868348) (0.425,0.00583490768403) (0.43,0.00776145763573) (0.435,0.00972468521334) (0.44,0.011728081332) (0.445,0.0137750917704) (0.45,0.0158689565792)}
\def\asymSweepCTwo{(0.01,0.18226248587) (0.015,0.171555572246) (0.02,0.161078717989) (0.025,0.150824512134) (0.03,0.140777815349) (0.035,0.130927250378) (0.04,0.121269359137) (0.045,0.11180113385) (0.05,0.102520165443) (0.055,0.09342505881) (0.06,0.0845158423172) (0.065,0.0757943402092) (0.07,0.0672644602798) (0.075,0.0589323174063) (0.08,0.0508060510683) (0.085,0.0461545258016) (0.09,0.0417475887831) (0.095,0.0374788168134) (0.1,0.0333582867036) (0.105,0.0293944214261) (0.11,0.0255924086959) (0.115,0.0219528701449) (0.12,0.0184708205201) (0.125,0.0151361216293) (0.13,0.0119403351473) (0.135,0.00888313531542) (0.14,0.00750643052502) (0.145,0.00676907508112) (0.15,0.00609473658629) (0.155,0.00547620778209) (0.16,0.00490513972996) (0.165,0.00437260192097) (0.17,0.00387005950945) (0.175,0.00339065425547) (0.18,0.00293026736731) (0.185,0.00248744080303) (0.19,0.00206197671654) (0.195,0.00165383719639) (0.2,0.00126291327265) (0.205,0.000888929987332) (0.21,0.000531288819329) (0.215,0.000188804743833) (0.22,0) (0.225,0) (0.23,0) (0.235,0) (0.24,0) (0.245,0) (0.25,0) (0.255,0) (0.26,0) (0.265,0) (0.27,0) (0.275,0) (0.28,0) (0.285,0) (0.29,0) (0.295,0) (0.3,0) (0.305,0) (0.31,0) (0.315,0) (0.32,0) (0.325,0) (0.33,0) (0.335,0) (0.34,0) (0.345,0) (0.35,0) (0.355,0) (0.36,0) (0.365,0) (0.37,0) (0.375,0) (0.38,0) (0.385,0) (0.39,0) (0.395,0) (0.4,0) (0.405,0) (0.41,0) (0.415,0) (0.42,0) (0.425,0) (0.43,0) (0.435,0) (0.44,0) (0.445,0) (0.45,0)}

\def\ctsTruckContinuous{(0.0,0.43070783095) (0.01,0.426433989364) (0.05,0.36612114233) (0.1,0.220796129198)}
\def\ctsTruckBinary{(0.0,0.43070783095) (0.01,0.426713687866) (0.05,0.388030661812) (0.1,0.238808942676)}
\def\ctsTruckErrors{(0.0,0.43070783095) +- (0,0.010246245145) (0.01,0.426433989364) +- (0,0.0103800479064) (0.05,0.36612114233) +- (0,0.010588652462) (0.1,0.220796129198) +- (0,0.010865955834)}
\def\ctsJointContinuous{(0.0,0.43070783095) (0.01,0.426353350136) (0.05,0.326444115934) (0.1,0.168622659555)}
\def\ctsJointBinary{(0.0,0.43070783095) (0.01,0.428978177552) (0.05,0.321679883237) (0.1,0.189823493788)}
\def\ctsJointErrors{(0.0,0.43070783095) +- (0,0.010246245145) (0.01,0.426353350136) +- (0,0.0101137690656) (0.05,0.326444115934) +- (0,0.0121222871116) (0.1,0.168622659555) +- (0,0.0125509243307)}

\def\gainBoundFullTieExact{(0.2,-0.0686359510824) (0.215,-0.0612583768905) (0.23,-0.0538321651491) (0.245,-0.0463852276875) (0.26,-0.0389543465924) (0.275,-0.0315867153062) (0.29,-0.0243407892025) (0.305,-0.0172862265003) (0.32,-0.0105027862057) (0.335,-0.00407816133284) (0.35,0.00189515418919) (0.365,0.00732375508397) (0.38,0.0121176730725) (0.395,0.0161948325199) (0.41,0.0194852474311) (0.425,0.021934618514) (0.44,0.0235070479656) (0.45,0.0240594653327) (0.455,0.0241866743834) (0.47,0.0239781301025) (0.485,0.0229058269683) (0.5,0.0210121733247) (0.515,0.0183549058214) (0.53,0.0150037779758) (0.545,0.0110368796198) (0.56,0.00653686671891) (0.575,0.00158736158786) (0.59,-0.00373025651716) (0.605,-0.00933950524954) (0.62,-0.0151707951813) (0.635,-0.021162729908) (0.65,-0.0272627315572) (0.665,-0.033427062485) (0.68,-0.0396203500243) (0.695,-0.0458147420765) (0.71,-0.0519888271041) (0.725,-0.0581264442633) (0.74,-0.0642154908162) (0.755,-0.0702468081727) (0.77,-0.0762131988009) (0.785,-0.0821085973828) (0.8,-0.0879273938664)}
\def\gainBoundFullTieLower{(0.2,-0.296741441634) (0.215,-0.275000558228) (0.23,-0.25357790831) (0.245,-0.23247349188) (0.26,-0.211687308938) (0.275,-0.191219359483) (0.29,-0.171069643516) (0.305,-0.151238161038) (0.32,-0.131724912046) (0.335,-0.112529896543) (0.35,-0.0936531145274) (0.365,-0.0750945659997) (0.38,-0.0568542509597) (0.395,-0.0389321694075) (0.41,-0.0213283213431) (0.425,-0.00404270676653) (0.44,0.0129246743223) (0.45,0.0240594653327) (0.455,0.0192169349941) (0.47,0.00447718831988) (0.485,-0.0105807918422) (0.5,-0.025957005492) (0.515,-0.0416514526296) (0.53,-0.0576641332551) (0.545,-0.0739950473683) (0.56,-0.0906441949693) (0.575,-0.107611576058) (0.59,-0.124897190635) (0.605,-0.142501038699) (0.62,-0.160423120251) (0.635,-0.178663435291) (0.65,-0.197221983819) (0.665,-0.216098765835) (0.68,-0.235293781338) (0.695,-0.254807030329) (0.71,-0.274638512808) (0.725,-0.294788228775) (0.74,-0.315256178229) (0.755,-0.336042361172) (0.77,-0.357146777602) (0.785,-0.37856942752) (0.8,-0.400310310925)}
\def\gainBoundFullTieUpper{(0.2,0.221102904824) (0.215,0.211773127442) (0.23,0.202125116573) (0.245,0.192158872215) (0.26,0.18187439437) (0.275,0.171271683037) (0.29,0.160350738217) (0.305,0.149111559908) (0.32,0.137554148112) (0.335,0.125678502828) (0.35,0.113484624056) (0.365,0.100972511796) (0.38,0.0881421660485) (0.395,0.0749935868132) (0.41,0.0615267740901) (0.425,0.0477417278793) (0.44,0.0336384481806) (0.45,0.0240594653327) (0.455,0.0295738219233) (0.47,0.0459047360365) (0.485,0.0619174166619) (0.5,0.0776118637996) (0.515,0.0929880774494) (0.53,0.108046057611) (0.545,0.122785804286) (0.56,0.137207317472) (0.575,0.151310597171) (0.59,0.165095643382) (0.605,0.178562456105) (0.62,0.19171103534) (0.635,0.204541381088) (0.65,0.217053493347) (0.665,0.229247372119) (0.68,0.241123017403) (0.695,0.2526804292) (0.71,0.263919607508) (0.725,0.274840552329) (0.74,0.285443263662) (0.755,0.295727741507) (0.77,0.305693985864) (0.785,0.315341996734) (0.8,0.324671774116)}
\def\gainBoundFullEqualityPoint{(0.45,0.0240594653327)}

\IEEEaftertitletext{\vspace{-12pt}}

\begin{document}
\maketitle
\acresetall
\begin{abstract}
In mobile wireless links, motion changes path directions and visibility,
and blockage can shift communication to a different propagation path.
\Ac{isac} allows a \ac{bs}
to use echoes of data transmissions to monitor these changes and guide
subsequent beam selection.
\commentadd{\reviewfocus{Illuminating objects that govern future path availability may reduce the beamforming gain on the current communication route.}}
Existing \commentadd{sensing--communication tradeoffs} and dynamic beam-control frameworks do not directly quantify the future communication gain from this \commentadd{current-rate sacrifice}. We quantify sensing value by the future-rate gain enabled by echo-adaptive beam selection. \commentadd{We derive explicit expressions for the future communication gain and the net gain after accounting for the current-rate loss.} The analysis uses a coupled communication--echo channel with binary-state Markov motion over a finite burst informed by one processed echo. We show that monitoring is worthwhile when its future-rate gain exceeds the current loss, \commentadd{while more informative observations yield zero additional future rate if they cannot change any future optimal beam.} We derive interpretable gain bounds exposing the roles of \commentadd{\reviewfocus{improved scene discrimination}}, motion memory and future beam-rate differences. \commentadd{Reciprocal-channel tests demonstrate positive net accumulated-rate gains from monitoring a moving blocker during reflected-path service, including continuous position uncertainty.}
\end{abstract}
\acresetall

\section{Introduction}
In mobile wireless links, motion changes the directions and visibility of
propagation paths. Blockage can weaken a previously useful path, while
another path becomes preferable for communication. \Ac{isac} allows a \ac{bs} to use echoes of its data
transmission to monitor these changes and guide subsequent beam selection.
\commentadd{\reviewfocus{The shared waveform creates a spatial tradeoff: a data beam
that illuminates objects governing future path availability may provide
less gain along the propagation route currently serving the \commentadd{\ac{ue}}.
}}
The design question is how much future rate this information recovers
and whether it compensates for the current loss.

Sensing-assisted methods handle changing paths through predictive
beamforming and sensing--communication stage optimization~\cite{du},
geometric object inference~\cite{ruanisac}, and
blockage-aware prediction with backup-path selection~\cite{cheng,ruanspawc}.
These approaches use scene information to anticipate motion and maintain
communication when useful paths change.
Complementary analytical frameworks evaluate the resulting decisions:
partially observable control jointly considers sensing and cumulative rate~\cite{hussain,ronquillo}, while
network-level analysis quantifies coverage and beam misalignment under
mobility and blockage~\cite{huangmanagement}. However, these methods and analyses do not directly quantify how much
current-rate loss a monitoring transmission can repay through later beam
selection.

\commentadd{In information-theoretic studies, the communication cost of
informative sensing is characterized by}
capacity--distortion~\cite{ahmadipour,xiong}. \commentadd{For future beam selection, motion can change the usefulness
of information acquired earlier. Open-loop Markov-state filtering~\cite{lindstrom}
and history-dependent signaling for unifilar channels~\cite{nikbakht}
account for temporal state dependence.}
Closer to path monitoring, communication--change-detection theory bounds
the rate versus detection-delay tradeoff within a specified class of codes
\cite{seo_qcd}. Its data channel is state-independent, whereas path
visibility here affects both communication and echoes.
However, these estimation and detection bounds do not directly quantify
how much future communication rate \commentadd{a more informative echo observation} can recover, or whether
that gain offsets the current-rate loss.

We quantify sensing value through its future communication-rate gain and the net gain after the current-rate loss. A binary-state Markov model connects the initial echo to later beam choices in coupled communication and echo channels. A finite service burst without later informative feedback permits explicit evaluation.
Our contributions are: \textit{i)} We derive explicit expressions for the future-rate gain and net gain
for comparing initial beams. This determines when to
monitor rather than maximize current rate and \commentadd{identifies when more informative observations have zero
additional future-rate gain despite retained motion information.}
\textit{ii)} We establish interpretable gain bounds exposing the roles of
\commentadd{\reviewfocus{improved scene discrimination}}, motion memory and future beam-rate differences.
\textit{iii)} We demonstrate positive net accumulated-rate gains through
reciprocal-channel tests with continuous-valued vehicle and blocker position uncertainty.

\section{System Model and Decision Problem}
\label{sec:model}
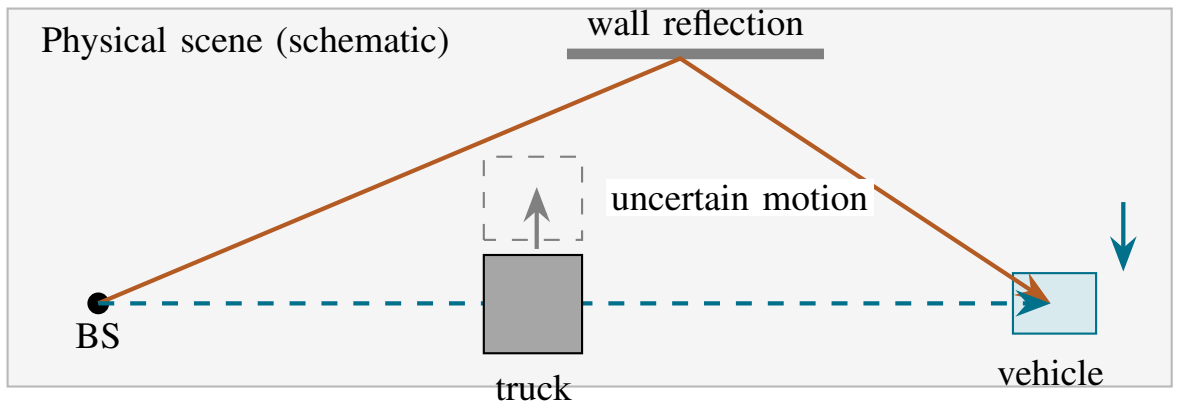
\begin{figure}[!t]
\centering
\begin{tikzpicture}[x=1cm,y=1cm,>=Stealth,font=\scriptsize]
\draw[gray!55,fill=gray!8] (-.1,0) rectangle (7.6,2.5);
\node[anchor=west] at (0,2.27) {Physical scene (schematic)};
\coordinate (bs) at (.5,.55);\coordinate (ue) at (6.8,.55);
\draw[fill=black] (bs) circle (.065);\node[below] at (bs){\acs{bs}};
\draw[fill=monitor!15,draw=monitor] (6.55,.35) rectangle (7.1,.75);
\node[below] at (6.8,.3){vehicle};
\draw[->,monitor,thick] (7.28,1.22)--(7.28,.76);
\draw[line width=2pt,gray] (3.6,2.2)--(5.3,2.2);
\node[above] at (4.45,2.2){wall reflection};
\draw[comm,thick,->] (bs)--(4.35,2.17)--(ue);
\draw[monitor,dashed,thick,->] (bs)--(ue);
\draw[fill=gray!70] (3.05,.22) rectangle (3.7,.87);
\node[below] at (3.38,.2){truck};
\draw[dashed,gray] (3.05,.97) rectangle (3.7,1.52);
\draw[->,gray,thick] (3.4,.91)--(3.4,1.32);
\node[anchor=west,fill=white,inner sep=1pt] at (3.85,1.25){uncertain motion};
\end{tikzpicture}
\caption{\commentadd{\acs{bs}--vehicle scene with direct and reflected paths.
A truck can block the direct route; while serving the \acs{ue} through
the reflection, the \acs{bs} may sacrifice current gain to monitor the
truck and predict LoS- recovery (Section~II-C).
Shared blockage affects communication and both echo legs.}}
\label{fig:scene}
\end{figure}

\commentadd{Fig.}~\ref{fig:scene} illustrates a \ac{bs} serving a vehicle as the
\ac{ue}: a moving truck can weaken the \ac{los} route and make a reflection
preferable. The \ac{bs} transmits data while illuminating the scene, uses
the resulting echo to update its scene estimate, and predicts path changes
to choose later communication beams. The current beam must therefore
balance the rate delivered now with the sensing information needed to
support communication in future service slots.
To index these transmissions,
let $i$ denote the current slot and let
$i+\delta_1,\ldots,i+\delta_L$ denote the $L$ future service slots, with
$\delta_\ell>0$. We use $\ell=0,\ldots,L$ to number the current and future
services, setting $\delta_0=0$.

To quantify the initial transmission's communication benefit exactly,
we use the following assumptions:

\begin{assumption}[Short-burst information and separable services]
\label{ass:short-burst}
The initial echo is processed before every future service; no later echo
or other informative feedback is available before the last decision.
Services have fixed feasible beam codebooks and separate power constraints,
with no shared resource constraint. Beams do not affect scene motion.
\end{assumption}

\vspace{-0.15cm}
\subsection{\commentadd{Communication and Sensing Models}}
The scene $S_\ell$ contains vehicle/blocker poses and propagation
coefficients. It determines the downlink $\mathbf h_c(S_\ell)\in\mathbb C^{N_T}$
and echo channel $\mathbf G(S_\ell)\in\mathbb C^{N_R\times N_T}$, where
$N_T,N_R$ are the numbers of transmit and receive antennas.
\commentadd{\reviewfocus{At service $\ell$, the \commentadd{\ac{bs}} selects a transmit beam from a
finite candidate family indexed by $\mathcal A_\ell$. For each index
$a\in\mathcal A_\ell$, the unit-norm vector
$\mathbf b_{i+\delta_\ell,a}$ specifies its antenna weights and hence its
direction and beam pattern. We write $\mathbf b_{\ell,a}$ for brevity.}}
\commentadd{\reviewfocus{Each service slot transmits $M$ symbols $U_{\ell,j}$,
$j=1,\ldots,M$, with $\E|U_{\ell,j}|^2=1$, using one beam and
constant scene channels.} The transmitted vector
$\mathbf x_{\ell,j}=\sqrt P\mathbf b_{\ell,a}U_{\ell,j}$ has average
power $P$. The received data signal at the \ac{ue} and echo signal at
the \ac{bs} are, respectively,}
\begin{align}
Y_{\ell,j}&=\sqrt P\,\mathbf h_c(S_\ell)^H\mathbf b_{\ell,a}U_{\ell,j}+W_{c,\ell,j},\notag\\
\mathbf Z_{\ell,j}&=\sqrt P\,\mathbf G(S_\ell)\mathbf b_{\ell,a}U_{\ell,j}
 +\mathbf W_{s,\ell,j}.
\label{eq:signal}
\end{align}
Here noises $W_c\sim\CN(0,2N_c)$ and $\mathbf W_s\sim\CN(\mathbf0,N_s\mathbf I)$
are independent across symbols/services. \commentadd{The \ac{ue} knows its effective channel for coherent decoding.}

\subsection{Scene uncertainty and motion prediction}
\commentadd{\reviewfocus{The scene $S_\ell$ contains continuous vehicle and blocker
positions and velocities whose exact values are unknown to the \ac{bs}.
To obtain an explicit analysis, we approximate this uncertainty by two
representative physical configurations $S_{\ell,+},S_{\ell,-}$ at each
service. Each specifies poses and channel coefficients as in Section~II-A.
The binary state $B_\ell\in\{+,-\}$ indicates whether the scene is
$S_{\ell,+}$ or $S_{\ell,-}$. The \ac{bs} knows the candidate configurations and channels but must
infer $B_\ell$.}

\reviewfocus{For example, near a blockage boundary in Fig.~\ref{fig:scene}, partition
the predicted pose distribution into configurations with an unblocked
or blocked \ac{los} path. Choose a feasible, high-probability pose per region as
$S_{\ell,+}$ or $S_{\ell,-}$.}

Configurations vary between slots, so $+$ and $-$ need not permanently
mean clear and blocked. We assume a binary Markov model with known,
possibly time-varying transitions $\Prb(B_{\ell+1}=s'\mid B_\ell=s)$,
$s,s'\in\{+,-\}$. The configurations and transitions are supplied by
a maintained motion model. Given $B_0$, future states are independent
of the initial echo; motion and echo noises are independent.

This is a local approximation to continuous scene uncertainty, useful
when within-region rate and echo-mean variation is small compared with
the contrast between representatives, such as that caused by an unblocked
versus blocked \ac{los} path. Decision-relevant state abstraction is also used
in~\cite{hussain}, which models mobility through strongest-beam indices
and blockage states. Section~\ref{sec:continuous} tests
its approximation using continuous positions and their mixture likelihoods.

Let $\mathcal H_i$ be the pretransmission \ac{bs} history and
$p=\Prb(B_0=+\mid\mathcal H_i)\in(0,1)$ the probability of the initial
$+$ configuration before the new echo. All probabilities and expectations
below condition on this fixed history. Composing the motion transitions
gives $q_{\ell,+}=\Prb(B_\ell=+\mid B_0=+)$ and
$q_{\ell,-}=\Prb(B_\ell=+\mid B_0=-)$: the two predictions of the future
$+$ label conditional on the initial label. Let $u\in[0,1]$ denote the
posterior probability of $B_0=+$ after observing the initial echo, and
let $q_\ell(u)$ denote the resulting predicted probability of $B_\ell=+$.
By total probability,
\begin{equation}
 q_\ell(u)=u q_{\ell,+}+(1-u)q_{\ell,-}=\alpha_\ell+\rho_\ell u,
 \label{eq:motion}
\end{equation}
where
\begin{equation}
 \alpha_\ell:=q_{\ell,-},\qquad
 \rho_\ell:=q_{\ell,+}-q_{\ell,-}.
 \label{eq:motion-parameters}
\end{equation}}

\subsection{Current rate and echo quality of the same beam}
\commentadd{\reviewfocus{We measure communication performance by the conditional
achievable rate $r(S,\mathbf b)$ in bit/symbol, given by
\begin{equation}
 r_{\rm G}(S,\mathbf b)=\log_2\!\left(
 1+P|\mathbf h_c(S)^H\mathbf b|^2/(2N_c)\right).
 \label{eq:gaussian-rate}
\end{equation}}
At service $\ell$, the rate in scene $s$ with beam $a$, and the
initial expected rate, are respectively
\begin{align}
 r_{\ell,s,a}&:=r(S_{\ell,s},\mathbf b_{\ell,a}),\qquad s\in\{+,-\},
 \label{eq:conditional-rate}\\
 C_a&=p r_{0,+,a}+(1-p)r_{0,-,a}.
 \label{eq:initial-rate}
\end{align}
\reviewfocus{For sensing, the \ac{bs} infers $B_\ell$ from the echoes
in~\eqref{eq:signal}. Matched filtering with the known transmitted symbols
gives the sufficient statistic (conditioned on those symbols and $B_\ell=s$)
\begin{equation}
 \overline{\mathbf Z}_{\ell,a}
 =\frac{\sum_{j=1}^{M}U_{\ell,j}^{*}\mathbf Z_{\ell,j}}{E_\ell}
 \sim\CN\!\left(\sqrt P\mathbf G(S_{\ell,s})\mathbf b_{\ell,a},
                 (N_s/E_\ell)\mathbf I\right).
 \label{eq:average}
\end{equation}}
where $E_\ell:=\sum_{j=1}^{M}|U_{\ell,j}|^2$ is the known block energy.
The data symbols are independent
of the scene.}
\commentadd{\reviewfocus{For the initial observation ($\ell=0$), projecting
\eqref{eq:average} along the difference between its two means and
normalizing the noise gives the equivalent scalar Gaussian echo model
\begin{equation}
 \widetilde Z_a\mid B_0=\pm\sim\mathcal N(\pm\sqrt{I_a},1).
 \label{eq:scalar-echo}
\end{equation}}
The associated detection \commentadd{\ac{snr}} is}
\commentadd{\begin{equation}
 I_a=\frac{PE_0}{2N_s}
 \|[\mathbf G(S_{0,+})-\mathbf G(S_{0,-})]\mathbf b_{0,a}\|^2.
 \label{eq:info}
\end{equation}}
The detection \ac{snr} $I_a$ determines how reliably the echo distinguishes
the two scenes, thereby affecting the scene posterior, future beam choices
and their communication rates.

\commentadd{\reviewfocus{Changing the initial beam $a$ can increase the echo detection \ac{snr}
$I_a$ in~\eqref{eq:info} while reducing the current expected communication
rate $C_a$ in~\eqref{eq:initial-rate}.} In Fig.~\ref{fig:scene}, for example,
the \commentadd{\ac{bs}} currently communicates through a reflection while monitoring a
moving blocker to predict future \commentadd{\ac{los}} recovery. Beam $\mathbf b_{0,c}$
points toward the wall reflection to serve the \commentadd{\ac{ue}}; beam $\mathbf b_{0,m}$
points toward the truck to monitor its motion. Different truck positions imply different probabilities of future
\ac{los} recovery. A dominant-direction approximation gives\( \mathbf h_c(S_{0,s})=g_s\mathbf b_{0,c},\)
 \(\mathbf G(S_{0,+})-\mathbf G(S_{0,-})
       =\mathbf u\mathbf b_{0,m}^{H}\), 
where $g_s$ is the reflected-channel gain and $\mathbf u\ne\mathbf0$
is the receive-side echo difference. For nonaligned unit-norm beams,
$|\mathbf b_{0,c}^{H}\mathbf b_{0,m}|<1$, and at least one $g_s\ne0$,
a conditional rate strictly increasing with received \commentadd{\ac{snr}} gives, using
\eqref{eq:conditional-rate}--\eqref{eq:initial-rate} and~\eqref{eq:info},
\begin{equation}
\reviewfocus{ C_c>C_m,\qquad
 I_c=|\mathbf b_{0,m}^{H}\mathbf b_{0,c}|^2 I_m<I_m.}
 \label{eq:spatial-ci-order}
\end{equation}
Thus beam $c$ delivers higher current rate, whereas beam $m$ provides
better scene discrimination at the same $P$, $M$ and sample energy $E_0$, at current cost
$C_c-C_m$. }

\subsection{\commentadd{Accumulated Communication Rate}}
\commentadd{We evaluate an initial beam $a$ by its current rate $C_a$ and
the future rates obtained using its echo. \reviewfocus{To select future beams, the
\ac{bs} first combines prior $p$ with echo $\widetilde Z_a$
in~\eqref{eq:scalar-echo} to obtain the probability of $B_0=+$:
\begin{equation}
 P_{I_a}:=\Prb(B_0=+\mid\mathcal H_i,E_0,\widetilde Z_a).
 \label{eq:posterior}
\end{equation}}
\reviewfocus{For fixed $p$ and known $E_0$, Bayes' rule and~\eqref{eq:scalar-echo}
give, with $\logit(u):=\ln(u/(1-u))$,
\begin{equation}
\begin{aligned}
 \logit(P_{I_a})&=\logit(p)+2\sqrt{I_a}\widetilde Z_a,\\
 \logit(P_{I_a})\mid B_0=\pm
 &\sim\mathcal N\!\left(\logit(p)\pm2I_a,\,4I_a\right).
\end{aligned}
\label{eq:posterior-law}
\end{equation}}
For $I_a>0$, $P_{I_a}$ follows a two-component logistic-normal mixture
with weights $p,1-p$. At $I_a=0$, $P_{I_a}=p$. The distribution depends on beam $a$
only through $I_a$ in~\eqref{eq:info}. The motion
map~\eqref{eq:motion} converts it to the future probability
$q_\ell(P_{I_a})$. For any such probability $q$, beam $a$ has expected
rate $R_{\ell,a}(q)$; maximizing over $a\in\mathcal A_\ell$ gives $F_\ell(q)$:
\begingroup
\medmuskip=2mu \thickmuskip=3mu \thinmuskip=2mu
\begin{equation}
 R_{\ell,a}(q)=qr_{\ell,+,a}+(1-q)r_{\ell,-,a},\;
 F_\ell(q)=\max_a R_{\ell,a}(q).
\label{eq:envelope}
\end{equation}
\endgroup
A future policy $\pi_\ell$ maps the available history
$\mathcal H_{i+\delta_\ell}$ to a beam index in $\mathcal A_\ell$.
Optimizing these policies gives the accumulated rate, with nonnegative
service weights $\beta_\ell$:
\begin{equation}
\begin{aligned}
 J_a&=C_a+\sup_\pi\E\!\left[\sum_{\ell=1}^{L}\beta_\ell
 r_{\ell,B_\ell,\pi_\ell(\mathcal H_{i+\delta_\ell})}\right]\\
 &\overset{\text{Assump.~\ref{ass:short-burst}}}{=}
 C_a+\sum_{\ell=1}^{L}\beta_\ell\E[F_\ell(q_\ell(P_{I_a}))].
\end{aligned}
 \label{eq:objective}
\end{equation}
The second equality holds because future beams provide no new information
within the window and share no resource budget, so each can maximize its
own expected rate. Unlike estimation-distortion or detection-delay
objectives~\cite{ahmadipour,lindstrom,seo_qcd}, $J_a$ includes the current
communication cost and the rate recovered through future beam selection.

We choose the initial beam $a$ to maximize $J_a$. Additionally, we
impose the current-rate requirement $C_a\ge R_{\min}$.}

\section{An Explicit Rate-Sacrifice Threshold}
\label{sec:threshold}
{\color{black}
To choose an initial beam, compare the current-rate-maximizing beam
$c$ with a monitoring candidate $m$, where $I_m>I_c$ and the current-rate
loss is $d_C:=C_c-C_m\ge0$. We compute the accumulated-rate difference
$J_m-J_c$: a positive value favors $m$, a negative value favors $c$,
and zero means a tie. Using~\eqref{eq:objective} under
Assumption~\ref{ass:short-burst}, this difference is
\begin{equation}
\begin{aligned}
 &J_m-J_c=-d_C\\[-2pt]
 &\quad+\sum_{\ell=1}^{L}\beta_\ell\Big\{
 \E[F_\ell(q_\ell(P_{I_m}))]
 -\E[F_\ell(q_\ell(P_{I_c}))]\Big\}.
\end{aligned}
 \label{eq:net-rate-start}
\end{equation}
\reviewfocus{We evaluate~\eqref{eq:net-rate-start} in three steps:
\textit{i)} write $F_\ell(q)$ using the switching probabilities
$\theta_{\ell,v}$ and slope increases $m_{\ell,v}$;
\textit{ii)} substitute $q=q_\ell(P_{I_a})$ to express future rates in
terms of the echo-updated initial-state probability $P_{I_a}$;
\textit{iii)} evaluate $\E[F_\ell(q_\ell(P_{I_a}))]$ as a function of
$I_a$, then substitute $I_m,I_c$ into~\eqref{eq:net-rate-start} to obtain
$J_m-J_c$.}

\textit{i) Rate $F_\ell(q)$ and its switching probabilities.}
From~\eqref{eq:envelope}, each candidate beam has expected rate
\[
 R_{\ell,a}(q)=r_{\ell,-,a}
       +q\big(r_{\ell,+,a}-r_{\ell,-,a}\big),
\]
which is affine in $q$: each beam gives a straight rate line with
intercept $r_{\ell,-,a}$ and slope $r_{\ell,+,a}-r_{\ell,-,a}$.
The best expected rate $F_\ell(q)=\max_{a\in\mathcal A_\ell}R_{\ell,a}(q)$
follows the highest of these lines and is continuous, piecewise linear
and convex.
Denote the probabilities where the maximizing beam changes by
$\theta_{\ell,1}<\theta_{\ell,2}<\cdots$ in $(0,1)$.
These beam-selection thresholds are the bends in $F_\ell$ shown in
Fig.~\ref{fig:rate-switch}; between successive thresholds the optimal
beam is unchanged.

\begin{figure}[!t]
\centering
\begin{tikzpicture}[x=5.2cm,y=3.6cm,>=Stealth,font=\scriptsize]
\draw[->] (0,.15)--(1.14,.15) node[right] {$q$};
\draw[->] (0,.15)--(0,.91);
\node[rotate=90] at (-.10,.53) {Expected rate at service $\ell$};
\draw[comm,dashed,thick] (0,.68)--(1,.28);
\draw[gray,dashed,thick] (0,.48)--(1,.58);
\draw[monitor,dashed,thick] (0,.2)--(1,.7);
\draw[black,very thick] (0,.68)--(.4,.52)--(.7,.55)--(1,.7);
\draw[gray,dotted] (.4,.15)--(.4,.52);
\draw[gray,dotted] (.7,.15)--(.7,.55);
\fill[black] (.4,.52) circle (1.4pt);
\fill[black] (.7,.55) circle (1.4pt);
\node[below] at (0,.15) {$0$};
\node[below] at (.4,.15) {$\theta_{\ell,1}$};
\node[below] at (.7,.15) {$\theta_{\ell,2}$};
\node[below] at (1,.15) {$1$};
\node[right,comm] at (1,.28) {$R_{\ell,a_1}(q)$};
\node[right,gray] at (1,.58) {$R_{\ell,a_2}(q)$};
\node[right,monitor] at (1,.7) {$R_{\ell,a_3}(q)$};
\node[above] at (.49,.8) {$F_\ell(q)$: highest rate};
\node at (.18,.01) {select $a_1$};
\node at (.55,.01) {select $a_2$};
\node at (.87,.01) {select $a_3$};
\end{tikzpicture}
\caption{Beam selection among three candidates. The envelope $F_\ell(q)$
follows the highest rate, switching from $a_1$ to $a_2$ at
$\theta_{\ell,1}$ and to $a_3$ at $\theta_{\ell,2}$.}
\label{fig:rate-switch}
\end{figure}
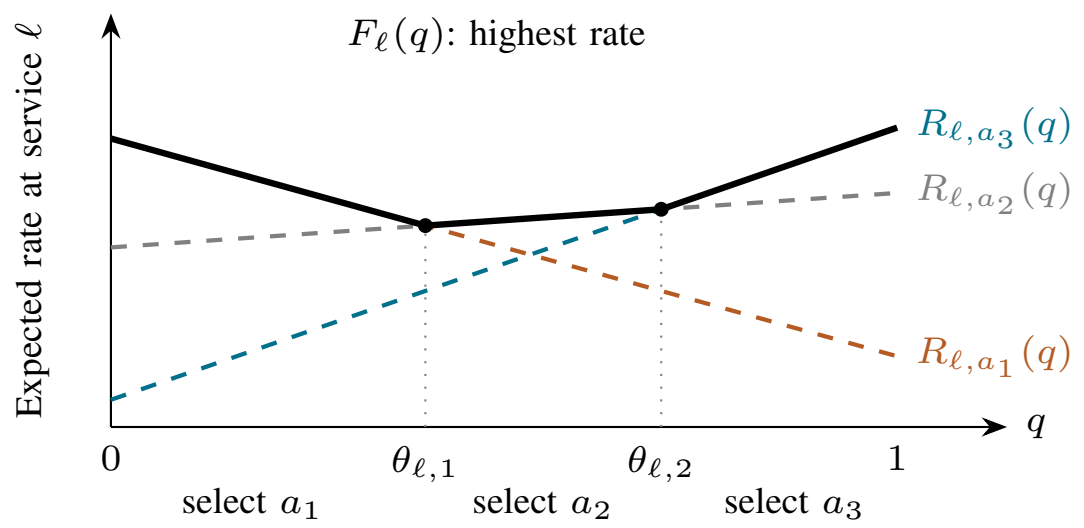

Index these switches by $v$. Let $F_\ell'(q)=\mathrm dF_\ell(q)/\mathrm dq$
and $m_{\ell,v}:=F_\ell'(\theta_{\ell,v}^{+})-
F_\ell'(\theta_{\ell,v}^{-})>0$, the increase from the left slope to the
right slope at switch $v$. With initial slope
$s_{\ell,0}:=F_\ell'(0^+)$ and $(z)_+:=\max(z,0)$, the rate envelope is
\begin{equation}
 F_\ell(q)=F_\ell(0)+s_{\ell,0}q
       +\sum_v m_{\ell,v}(q-\theta_{\ell,v})_+.
 \label{eq:multi-beam-hinge}
\end{equation}

\label{sec:echo-substitution}
\textit{ii) From $q$ to $P_{I_a}$.}
\reviewfocus{Equation~\eqref{eq:multi-beam-hinge} expresses the future rate as a
function of the scene probability $q$; we need its expectation as a
function of echo quality $I_a$. We therefore substitute
$q=q_\ell(P_{I_a})$ using the motion map~\eqref{eq:motion}, where the
distribution of $P_{I_a}$ is determined by $I_a$.} Taking expectations
in~\eqref{eq:multi-beam-hinge} gives
\begin{equation}
\begin{aligned}
 \E[F_\ell(q_\ell(P_{I_a}))]
 &=F_\ell(0)+s_{\ell,0}\E[q_\ell(P_{I_a})]\\
 &\quad+\sum_v m_{\ell,v}\E[(q_\ell(P_{I_a})-\theta_{\ell,v})_+].
\end{aligned}
\label{eq:rate-average-decomposition}
\end{equation}
\reviewfocus{To compute $J_m-J_c$ from~\eqref{eq:rate-average-decomposition},
we discard terms identical for $m$ and $c$. First, averaging over echoes
recovers the initial prior, $\E[P_{I_a}]=p$, so
$\E[q_\ell(P_{I_a})]=q_\ell(p)$. The common affine term
$F_\ell(0)+s_{\ell,0}q_\ell(p)$ therefore cancels.
Second, since $P_{I_a}\in[0,1]$, motion confines $q_\ell(P_{I_a})$
to the interval between $q_{\ell,+}$ and $q_{\ell,-}$.} Thresholds above or below this interval contribute $0$ or
$q_\ell(p)-\theta_{\ell,v}$, respectively, independently of echo quality. Such terms cancel between $m$ and $c$, including at the
endpoints. Only terms whose thresholds lie strictly inside this interval
can remain in $J_m-J_c$; let $\mathcal V_\ell$ collect their threshold indices:
\begin{equation}
 \min(q_{\ell,+},q_{\ell,-})<\theta_{\ell,v}
 <\max(q_{\ell,+},q_{\ell,-}).
 \label{eq:reachable}
\end{equation}
Only these thresholds contribute to the net-rate difference:
\begin{equation}
\begin{aligned}
 &J_m-J_c=-d_C+\sum_{\ell=1}^{L}\beta_\ell
              \sum_{v\in\mathcal V_\ell}m_{\ell,v}\\
 &\quad\times\big\{\E[(q_\ell(P_{I_m})-\theta_{\ell,v})_+]
 -\E[(q_\ell(P_{I_c})-\theta_{\ell,v})_+]\big\}.
\end{aligned}
\label{eq:net-reachable}
\end{equation}
Using $q_\ell(P_{I_a})=\alpha_\ell+\rho_\ell P_{I_a}$
from~\eqref{eq:motion}, with coefficients defined
in~\eqref{eq:motion-parameters}, \eqref{eq:net-reachable} becomes
\begin{equation}
\begin{aligned}
 &J_m-J_c=-d_C+\sum_{\ell=1}^{L}\beta_\ell
              \sum_{v\in\mathcal V_\ell}\kappa_{\ell,v}\\
 &\quad\times\big\{\E[(P_{I_m}-t_{\ell,v})_+]
             -\E[(P_{I_c}-t_{\ell,v})_+]\big\},
\end{aligned}
\label{eq:net-posterior}
\end{equation}
with initial-posterior threshold and rate weight
\begin{equation}
 t_{\ell,v}:=({\theta_{\ell,v}-\alpha_\ell})/{\rho_\ell},
 \qquad \kappa_{\ell,v}:=m_{\ell,v}|\rho_\ell|.
 \label{eq:initial-threshold}
\end{equation}
Here $q_\ell(t_{\ell,v})=\theta_{\ell,v}$ and $t_{\ell,v}\in(0,1)$.
The conversion holds for either sign of $\rho_\ell$, using
$(t-u)_+=(u-t)_++t-u$ and $\E[P_{I_m}]=\E[P_{I_c}]=p$. If $\rho_\ell=0$, $\mathcal V_\ell$ is empty.

\textit{iii) From echo quality $I_a$ to $J_m-J_c$.}
\reviewfocus{For detection \ac{snr} $I$, the posterior $P_I$ follows the
two-component logistic-normal mixture in~\eqref{eq:posterior-law}.
The positive-part expectation in~\eqref{eq:net-posterior} is}
\begin{align}
 A(p,I,t)&:=\E[(P_I-t)_+]\notag\\
 &=(1-t)p\Phi\!\left(\frac{2I-d}{2\sqrt I}\right)
 -t(1-p)\Phi\!\left(\frac{-2I-d}{2\sqrt I}\right),\notag\\
 d&=\logit(t)-\logit(p).
 \label{eq:hinge}
\end{align}
Here $\Phi$ is the standard normal \ac{cdf}, and
$A(p,0,t)=(p-t)_+$ by continuity.

Substituting~\eqref{eq:hinge} into~\eqref{eq:net-posterior} gives
each service's gain:
\begin{align}
 &\E[F_\ell(q_\ell(P_{I_m}))]-\E[F_\ell(q_\ell(P_{I_c}))]\notag\\
 &\quad=\sum_{v\in\mathcal V_\ell}\kappa_{\ell,v}
 [A(p,I_m,t_{\ell,v})-A(p,I_c,t_{\ell,v})].
 \label{eq:service-rate-difference}
\end{align}
Summing gives the future gain $d_C^\star$, compared with cost $d_C$.

\begin{theorem}[Maximum worthwhile current-rate sacrifice]
\label{thm:threshold}
For the initial beam pair $c,m$ above, assume the binary Markov scene
model of Section~\ref{sec:model}, the Gaussian echo~\eqref{eq:scalar-echo}
and Assumption~\ref{ass:short-burst}. Conditional on known sample energy
$E_0$, the accumulated future-rate gain for the rates
in~\eqref{eq:conditional-rate} is
\begin{equation}
 d_C^\star=\sum_{\ell,v}\beta_\ell\kappa_{\ell,v}
 [A(p,I_m,t_{\ell,v})-A(p,I_c,t_{\ell,v})],
 \label{eq:threshold}
\end{equation}
where the sum runs over services $\ell=1,\ldots,L$ and
indices $v\in\mathcal V_\ell$ from~\eqref{eq:reachable}, using
$t_{\ell,v},\kappa_{\ell,v}$ in~\eqref{eq:initial-threshold} and $A$
in~\eqref{eq:hinge}. For~\eqref{eq:objective},
$J_m-J_c=d_C^\star-d_C$. Monitoring is strictly preferable if and
only if $d_C<d_C^\star$. If no positively weighted service has a reachable
switch, improving the initial echo has zero future-rate value.
Furthermore,
\begin{equation}
 d_C^\star\le\sum_{\ell,v}\beta_\ell\kappa_{\ell,v}
 [p(1-t_{\ell,v})-A(p,I_c,t_{\ell,v})].
 \label{eq:ceiling}
\end{equation}
If this ceiling is at most $d_C$, even perfect initial sensing cannot
repay that current cost.
\end{theorem}
\begin{proof}
The identity $\E[P_I\mathbf1_{P_I>t}]=p\Prb(P_I>t\mid B_0=+)$
and Gaussian law~\eqref{eq:posterior-law} give~\eqref{eq:hinge}.
The gain formula~\eqref{eq:threshold} follows
from~\eqref{eq:service-rate-difference} and~\eqref{eq:net-rate-start}.
For $u\in[0,1]$, $(u-t)_+\le(1-t)u$ implies
$A(p,I,t)\le p(1-t)$, yielding~\eqref{eq:ceiling}.
\end{proof}

Theorem~\ref{thm:threshold} gives three design conclusions:
\textit{i) Beam selection.} Compare $d_C^\star$ with $d_C$ to choose
$m$ or $c$ (Fig.~\ref{fig:threshold-computation}); equality gives a tie.
\textit{ii) Information value.} Without a positively weighted reachable
switch, better sensing gives zero future-rate gain even if
$\rho_\ell\ne0$. With such a switch, $I_m>I_c$ gives positive future
gain, which must still exceed $d_C$.
\textit{iii) Cost rejection.} The ceiling~\eqref{eq:ceiling} uses only
future rates, motion, prior and $I_c$, avoiding the candidate's $I_m$
and \ac{cdf} evaluations. Costs at or above it cannot be repaid;
it is tight as $I_m\to\infty$ despite random future motion.

The explicit rate return distinguishes this result from the
mutual-information surrogate used for beam tracking~\cite{ronquillo}:
scene discrimination is valued through the future beam choices it changes.
Relative to open-loop estimation tradeoffs~\cite{lindstrom} and
state-independent-data detection tradeoffs~\cite{seo_qcd}, the result
evaluates echo-adaptive future communication in a scene affecting both
communication and sensing channels.

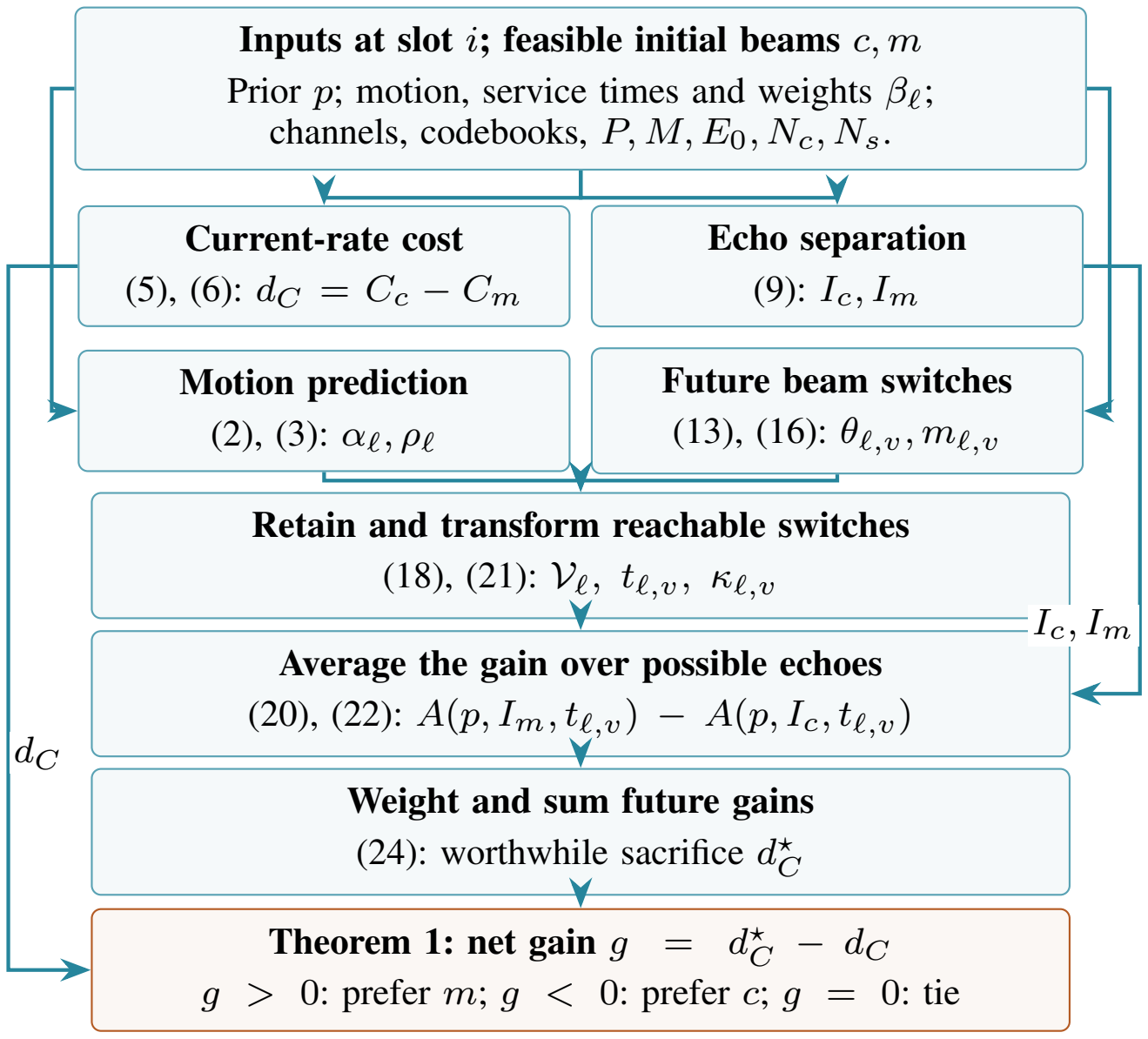
\begin{figure}[!t]
\centering
\begin{tikzpicture}[x=1cm,y=1cm,>=Stealth,font=\footnotesize,
 box/.style={draw=monitor!65,fill=monitor!4,rounded corners=2pt,
 align=center,inner sep=4pt},
 branch/.style={box,text width=3.45cm},
 flowstep/.style={box,text width=7.15cm},
 link/.style={->,thick,draw=monitor!85}]
\node[box,text width=7.4cm] (inputs) at (0,0) {
\textbf{Inputs at slot $i$; feasible initial beams $c,m$}\\[2pt]
Prior $p$; motion, service times and weights $\beta_\ell$;\\
channels, codebooks, $P,M,E_0,N_c,N_s$.};
\node[branch] (cost) at (-1.95,-1.35) {
\textbf{Current-rate cost}\\[2pt]
\eqref{eq:conditional-rate}, \eqref{eq:initial-rate}: $d_C=C_c-C_m$};
\node[branch] (echo) at (1.95,-1.35) {
\textbf{Echo separation}\\[2pt]
\eqref{eq:info}: $I_c,I_m$};
\draw[link] (inputs.south)--(0,-.83)-|(cost.north);
\draw[link] (inputs.south)--(0,-.83)-|(echo.north);
\node[branch] (motion) at (-1.95,-2.45) {
\textbf{Motion prediction}\\[2pt]
\eqref{eq:motion}, \eqref{eq:motion-parameters}: $\alpha_\ell,\rho_\ell$};
\node[branch] (beams) at (1.95,-2.45) {
\textbf{Future beam switches}\\[2pt]
\eqref{eq:envelope}, \eqref{eq:multi-beam-hinge}: $\theta_{\ell,v},m_{\ell,v}$};
\draw[link] (inputs.west)--(-4.02,0)|-(motion.west);
\draw[link] (inputs.east)--(4.02,0)|-(beams.east);
\node[flowstep] (filter) at (0,-3.55) {
\textbf{Retain and transform reachable switches}\\[2pt]
\eqref{eq:reachable}, \eqref{eq:initial-threshold}: $\mathcal V_\ell,\ t_{\ell,v},\ \kappa_{\ell,v}$};
\draw[link] (motion.south)--(-1.95,-2.98)-|(filter.north);
\draw[link] (beams.south)--(1.95,-2.98)-|(filter.north);
\node[flowstep] (average) at (0,-4.6) {
\textbf{Average the gain over possible echoes}\\[2pt]
\eqref{eq:net-posterior}, \eqref{eq:hinge}: $A(p,I_m,t_{\ell,v})-A(p,I_c,t_{\ell,v})$};
\draw[link] (filter)--(average);
\draw[link] (echo.east)--(4.25,-1.35)|-(average.east);
\node[anchor=east,fill=white,inner sep=1pt] at (4.25,-4.08) {$I_c,I_m$};
\node[flowstep] (sum) at (0,-5.65) {
\textbf{Weight and sum future gains}\\[2pt]
\eqref{eq:threshold}: worthwhile sacrifice $d_C^\star$};
\draw[link] (average)--(sum);
\node[flowstep,draw=comm,fill=comm!5] (decision) at (0,-6.7) {
\textbf{Theorem~\ref{thm:threshold}: net gain $g=d_C^\star-d_C$}\\[2pt]
$g>0$: prefer $m$; $g<0$: prefer $c$; $g=0$: tie};
\draw[link] (sum)--(decision);
\draw[link] (cost.west)--(-4.35,-1.35)|-(decision.west);
\node[anchor=west,fill=white,inner sep=1pt] at (-4.35,-5.05) {$d_C$};
\end{tikzpicture}
\caption{From known inputs to the net communication gain in
Theorem~\ref{thm:threshold}.}
\label{fig:threshold-computation}
\end{figure}

\subsection{Simpler Gain Bounds and Physical Implications}
Evaluating~\eqref{eq:threshold} requires $A$ at every reachable switch.
The following bounds replace these evaluations by one detection factor
and the distances from the prior $p$ to the initial-posterior thresholds
$t_{\ell,v}$. They give sufficient conditions for choosing or rejecting
monitoring without computing the exact gain.

\begin{corollary}[Asymmetric gain bounds and a condition for equality]
\label{cor:memory}
Under Theorem~\ref{thm:threshold}, with $0<p<1$, define
\begin{equation}
\begin{aligned}
 G_{\rm bal}&=2p(1-p)[\Phi(\sqrt{I_m})-\Phi(\sqrt{I_c})]
             \sum_{\ell,v}\beta_\ell\kappa_{\ell,v},\\
 \varepsilon_{\rm bal}&=\sum_{\ell,v}\beta_\ell\kappa_{\ell,v}
                                  |t_{\ell,v}-p|,
\end{aligned}
\label{eq:balanced-ingredients}
\end{equation}
where the sums retain $v\in\mathcal V_\ell$. Then
\begin{equation}
 G_{\rm bal}-\varepsilon_{\rm bal}-d_C
 \le J_m-J_c\le
 G_{\rm bal}+\varepsilon_{\rm bal}-d_C.
\label{eq:balanced-bound}
\end{equation}
Thus $d_C<G_{\rm bal}-\varepsilon_{\rm bal}$ guarantees positive net
gain, while $d_C\ge G_{\rm bal}+\varepsilon_{\rm bal}$ rules it out.

The bounds coincide if every positively weighted reachable threshold satisfies
$q_\ell(p)=\theta_{\ell,v}$, equivalently $t_{\ell,v}=p$, yielding
\begin{equation}
 J_m-J_c=G_{\rm bal}-d_C.
\label{eq:balanced-exact}
\end{equation}
Equality places $q_\ell(p)$ at each relevant beam-selection tie,
without requiring equal priors or symmetric rates or motion.

\end{corollary}
\begin{proof}
Let $D(t)=A(p,I_m,t)-A(p,I_c,t)$. Then
\[
 D(t)-D(p)=-\int_p^t
 [\Prb(P_{I_m}>z)-\Prb(P_{I_c}>z)]\,\mathrm dz.
\]
Thus $|D(t)-D(p)|\le|t-p|$. Setting $t=p$ in~\eqref{eq:hinge}
gives $D(p)=2p(1-p)[\Phi(\sqrt{I_m})-\Phi(\sqrt{I_c})]$.
Applying this to~\eqref{eq:threshold} proves the bounds, which
coincide when the mismatch is zero.
\end{proof}

In~\eqref{eq:balanced-bound}, $G_{\rm bal}$ is the rate gain when the
relevant beam choices are tied before the echo; $\varepsilon_{\rm bal}$
bounds the correction away from those ties. With $p$, motion, future rates
and current cost fixed, a larger detection improvement
$\Phi(\sqrt{I_m})-\Phi(\sqrt{I_c})$ raises $G_{\rm bal}$ and favors $m$;
a larger current cost $d_C$ favors $c$. Thresholds closer to $p$ reduce
the gap between the bounds. For two beams meeting at a switch,
$\beta_\ell\kappa_{\ell,v}|t_{\ell,v}-p|
=\beta_\ell m_{\ell,v}|q_\ell(p)-\theta_{\ell,v}|$ is their weighted
pre-echo expected-rate difference. If the bounds straddle zero,
Theorem~\ref{thm:threshold} gives the exact decision.
Motion scales each contribution through
$\kappa_{\ell,v}=m_{\ell,v}|\rho_\ell|$. For symmetric transitions with
fixed flip probability $s$, $|\rho_\ell|=|1-2s|^\ell$: motion near
$s=1/2$ suppresses the monitoring benefit, whereas near-deterministic
persistence or reversal retains information.

}

\section{Numerical Results}
\label{sec:results}
We test net gains, asymmetric bounds and continuous-position robustness.
Gains sum four equally weighted service rates in bit/symbol; divide by
four for the mean.

{\color{black}
\subsection{Reciprocal-geometry setup}
\label{sec:geometry-setup}
Co-located 16-element half-wavelength transmit/receive \acp{ula} at
$(0,0)$ serve a vehicle at $(10,y)$ m through direct and wall-reflected
paths, with reflector $(5,3)$ m and truck
$[4.8,5.2]\times[y_b-0.35,y_b+0.35]$ m. Across the four services,
$y=(0,0.15,0.30,0.45)$ m and $(y_{b,+},y_{b,-})$ are
$(0.20,0),(0.60,0.05),(0.72,0.10),(0.84,0.15)$ m.
Initially both configurations block \ac{los}; later, only $-$ does.
This implements reflection-based current service and truck monitoring
(Fig.~\ref{fig:scene}) with prescribed, rather than calibrated, motion transitions.

The 16 candidates combine eight coherent wall/second-direction mixtures
with coefficients $0,0.05,0.15,0.30,0.50,0.70,0.85,1$ and two tapers
$e^{-\omega d_n^2}$, $d_n=(n-7.5)/7.5$, $n=0,\ldots,15$,
$\omega\in\{0,8\}$. The second direction points toward the truck
initially and the vehicle later. Beams are normalized including
nonorthogonal cross terms; mixture coefficients are not power fractions.

Direct amplitudes are $0.05$ for truck-intersecting segments and $1$
otherwise; reflected amplitude is $0.45$. Two fixed-coefficient scatterers
per vehicle/truck generate reciprocal echo channels
$\mathbf G=\sum_q\gamma_q\mathbf l_q\mathbf l_q^T$, where
$\mathbf l_q$ combines direct and reflected propagation, including
phase at wavelength $0.05$ m. Intersection attenuation applies to both
legs, excluding terminal contact with a truck scatterer.
We use $P=2$, $M=128$, $N_c=0.3$ and $N_s=64\nu^2$.

\reviewfocus{Rates follow~\eqref{eq:gaussian-rate}; Gaussian payloads give
$E_0\sim\operatorname{Gamma}(M,1)$. Future policies know $E_0$;
initial objectives and bounds average over its common law before
payload generation.} Using 16 or 32 energy quadrature points changes results by less than $10^{-10}$.

\subsection{Asymmetric net gains, bounds, and equality}
We compare the current-rate-optimal initial beam $c$ with the design
maximizing~\eqref{eq:objective}. Both use the same 16 candidates and
current-rate floor $0.48$, then optimize future beams under their own
echoes. We sweep $p\in\{0.40,0.45,0.50\}$, $\nu\in\{0.5,1\}$ and
$s\in[0.01,0.45]$ with transition matrix
$\left(\begin{smallmatrix}1-2s&2s\\s&1-s\end{smallmatrix}\right)$.

Fig.~\ref{fig:asymmetricsweep}(a) demonstrates that monitoring can repay
its current cost under asymmetric motion. \reviewfocus{At $p=0.45$, $s=0.05$,
$\nu=0.5$, redirecting the wall beam toward the truck with an untapered
$0.30$ mixture sacrifices $0.15$ now but yields net gain $0.43$;
Corollary~\ref{cor:memory} certifies positive return.}
The advantage disappears near $s=1/3$, where identical transition rows
erase initial-state information, but can reappear under negative
correlation. Thus predictable reversal, like persistence, can justify
monitoring. Lower echo noise
benefits both choices, so their rate difference need not increase.

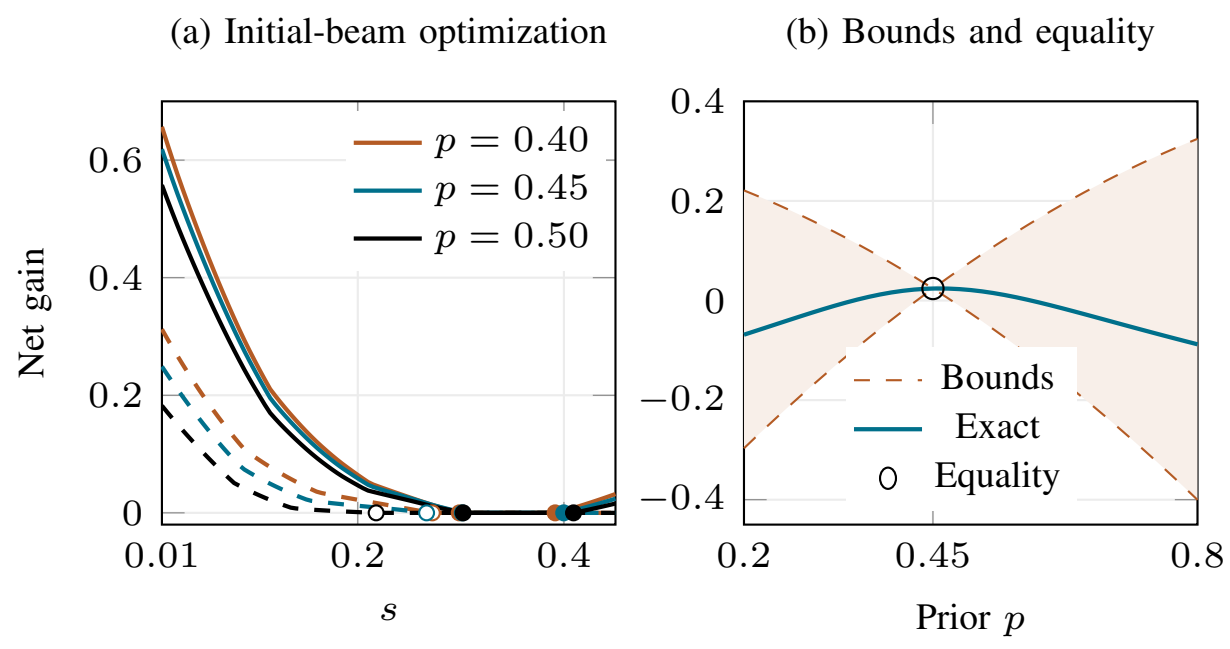
\begin{figure}[!t]
\centering
\begin{tikzpicture}
\pgfplotsset{compactresult/.style={scale only axis,width=3cm,height=2.8cm,
 title style={font=\scriptsize},grid=major,grid style={gray!15},scaled ticks=false,
 tick label style={font=\scriptsize,/pgf/number format/fixed,/pgf/number format/precision=2},
 label style={font=\scriptsize},
 legend style={font=\scriptsize,at={(.98,.98)},anchor=north east,
 draw=none,fill=white,inner sep=1pt,row sep=-1pt},
 legend image post style={xscale=.75}}}
\begin{axis}[compactresult,xmin=.01,xmax=.45,ymin=-.02,ymax=.70,
 title={(a) Initial-beam optimization},xlabel={$s$},ylabel={Net gain},
 xtick={.01,.2,.4},ytick={0,.2,.4,.6}]
\addplot[comm,thick] coordinates {\asymSweepAOne};\addlegendentry{$p=0.40$}
\addplot[comm,dashed,thick,forget plot] coordinates {\asymSweepATwo};
\addplot[monitor,thick] coordinates {\asymSweepBOne};\addlegendentry{$p=0.45$}
\addplot[monitor,dashed,thick,forget plot] coordinates {\asymSweepBTwo};
\addplot[black,thick] coordinates {\asymSweepCOne};\addlegendentry{$p=0.50$}
\addplot[black,dashed,thick,forget plot] coordinates {\asymSweepCTwo};
\addplot[comm,only marks,mark=*,mark size=1.4pt,mark options={fill=comm},forget plot] coordinates {(0.298455253231,0) (0.391483181943,0)};
\addplot[comm,only marks,mark=*,mark size=1.4pt,mark options={fill=white},forget plot] coordinates {(0.272330096085,0)};
\addplot[monitor,only marks,mark=*,mark size=1.4pt,mark options={fill=monitor},forget plot] coordinates {(0.301399472679,0) (0.399799147838,0)};
\addplot[monitor,only marks,mark=*,mark size=1.4pt,mark options={fill=white},forget plot] coordinates {(0.266742833388,0)};
\addplot[black,only marks,mark=*,mark size=1.4pt,mark options={fill=black},forget plot] coordinates {(0.302222681543,0) (0.409296922904,0)};
\addplot[black,only marks,mark=*,mark size=1.4pt,mark options={fill=white},forget plot] coordinates {(0.217843729872,0)};
\end{axis}
\begin{axis}[compactresult,at={(3.85cm,0)},anchor=south west,
 xmin=.2,xmax=.8,ymin=-.45,ymax=.40,
 title={(b) Bounds and equality},xlabel={Prior $p$},
 xtick={.2,.45,.8},ytick={-.4,-.2,0,.2,.4},
 yticklabel style={/pgf/number format/precision=2},
 legend style={at={(.48,.03)},anchor=south}]
\addplot[name path=fulltielower,comm,dashed] coordinates {\gainBoundFullTieLower};\addlegendentry{Bounds}
\addplot[name path=fulltieupper,comm,dashed,forget plot] coordinates {\gainBoundFullTieUpper};
\addplot[comm!10,forget plot] fill between[of=fulltielower and fulltieupper];
\addplot[monitor,thick] coordinates {\gainBoundFullTieExact};\addlegendentry{Exact}
\addplot[black,only marks,mark=o,mark size=2pt] coordinates {\gainBoundFullEqualityPoint};\addlegendentry{Equality}
\end{axis}
\end{tikzpicture}
\caption{\commentadd{Net gains including current cost; three future services,
16 beams.
(a) Optimal accumulated-rate design versus current-rate baseline:
solid $\nu=0.5$, dashed $\nu=1$; circles mark zero net gain.
(b) Fixed wall/$0.30$-mixture pair under a
separate time-varying transition model; bounds coincide at $p=0.45$.
Values average Gaussian payload sample energy.}}
\label{fig:asymmetricsweep}
\label{fig:gain-bounds}
\end{figure}

Fig.~\ref{fig:gain-bounds}(b) numerically verifies the equality condition
in Corollary~\ref{cor:memory}: the bounds coincide with the exact gain
at beam-selection ties and separate as threshold mismatch increases. For the wall/$0.30$-mixture pair, use
$q_\ell(u)=\theta_\ell+\rho_\ell(u-0.45)$, where $\theta_\ell$ is the
switch nearest $0.35$ and $(\rho_1,\rho_2,\rho_3)=(0.30,0.20,0.15)$.
These valid time-varying transitions leave one reachable switch per
service. At $p=0.45$, all relevant $t_{\ell,v}=p$, so both bounds recover
the positive net gain before and after energy averaging.

\subsection{Continuous-Position Validation}
\label{sec:continuous}
At $p=0.45$, $\nu=0.5$, $s=0.05$, we test the fixed wall/$0.30$-mixture
pair with persistent uniform offsets in $[-w,w]$ m for the truck alone
or independently vehicle and truck. Initial \ac{los} remains blocked;
labels retain the prescribed Markov dynamics.
The continuous controller integrates echo likelihoods and rates over
positions and labels; the binary controller uses nominal configurations.
Each selects among the same 16 future beams using realized sample energy.
Gains compare initial $m$ against $c$ for each controller on actual
continuous channels. We use 24,000 paired draws; visibility-split
quadrature refinement changes joint gains by less than $5\times10^{-5}$.

Fig.~\ref{fig:continuous} shows that positive monitoring gains survive
continuous-position uncertainty, while the nominal binary model can
overestimate their magnitude. Agreement at $w=0$ gives way to smaller
gains as phase and visibility vary within each configuration.
The binary controller's larger gain relative to its own baseline does
not imply higher throughput: the continuous controller achieves higher
absolute throughput after $m$ in the joint-uncertainty case at $w=0.10$ m.
Thus the representative model captures the benefit here, but its
accuracy depends on within-configuration variation.

\begin{figure}[!t]
\centering
\begin{tikzpicture}
\begin{axis}[name=truck,scale only axis,width=3cm,height=2.7cm,xmin=0,xmax=.1,ymin=0,ymax=.50,
 title={(a) Truck},title style={font=\footnotesize},
 xlabel={$w$ (m)},
 ylabel={Net accumulated gain},tick label style={font=\scriptsize,/pgf/number format/fixed,
 /pgf/number format/precision=2},
 label style={font=\scriptsize},grid=major,grid style={gray!15},
 xtick={0,.05,.1},ytick={0,.2,.4},
 legend style={font=\scriptsize,at={(.03,.04)},anchor=south west,
 draw=none,fill=white,inner sep=1pt,row sep=-1pt}]
\addplot[monitor,thick] coordinates {\ctsTruckContinuous};\addlegendentry{Continuous}
\addplot[comm,dashed,thick] coordinates {\ctsTruckBinary};\addlegendentry{Binary}
\addplot[black,dashdotted] coordinates {(0,0.430617328269)(.1,0.430617328269)};\addlegendentry{Nominal}
\addplot[monitor,only marks,mark=*,mark size=1.3pt,forget plot,error bars/.cd,
 y dir=both,y explicit] coordinates {\ctsTruckErrors};
\end{axis}
\begin{axis}[at={(truck.south east)},anchor=south west,xshift=0.85cm,
 scale only axis,width=3cm,height=2.7cm,xmin=0,xmax=.1,ymin=0,ymax=.50,
 title={(b) Vehicle and truck},title style={font=\footnotesize},
 xlabel={$w$ (m)},
 tick label style={font=\scriptsize,/pgf/number format/fixed,
 /pgf/number format/precision=2},label style={font=\scriptsize},
 grid=major,grid style={gray!15},xtick={0,.05,.1},ytick={0,.2,.4}]
\addplot[monitor,thick] coordinates {\ctsJointContinuous};
\addplot[comm,dashed,thick] coordinates {\ctsJointBinary};
\addplot[black,dashdotted] coordinates {(0,0.430617328269)(.1,0.430617328269)};
\addplot[monitor,only marks,mark=*,mark size=1.3pt,forget plot,error bars/.cd,
 y dir=both,y explicit] coordinates {\ctsJointErrors};
\end{axis}
\end{tikzpicture}
\caption{\commentadd{Continuous-position accumulated-rate difference $J_m-J_c$,
using the indicated controller for future beams. $w$: uncertainty half-width; bars: 95\% sampling intervals;
dash-dotted: nominal binary prediction.}}
\label{fig:continuous}
\end{figure}
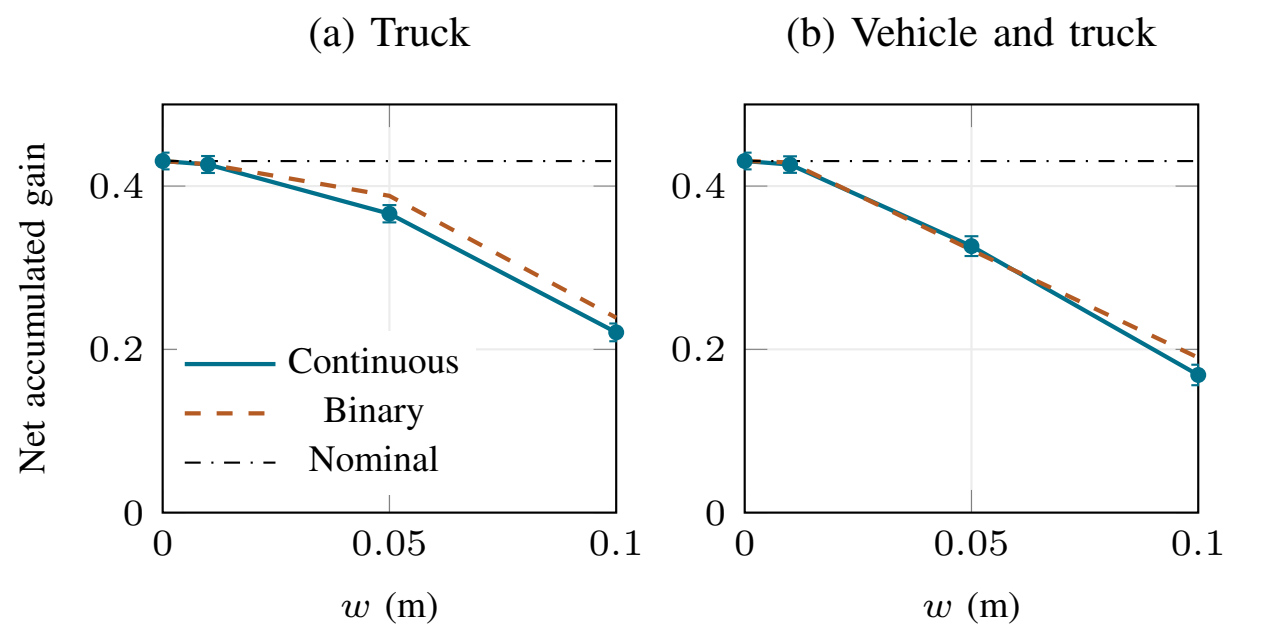

}

\section{Conclusion}

We evaluate sensing through the future communication-rate gain enabled by echo-adaptive beam selection. \commentadd{For general conditional achievable rates, binary-state Markov motion and a service burst informed by one processed echo, we derive this gain and its net gain explicitly.} \commentadd{More informative observations have zero additional future-rate value when they cannot change a future optimal beam;} otherwise its rate return must exceed the current loss. Interpretable bounds relate this value to \commentadd{\reviewfocus{improved scene discrimination}}, motion memory and future beam-rate differences. Reciprocal-channel experiments show positive net gains under continuous position uncertainty and expose the representative-state approximation's limits.

\end{document}